\documentclass[fleqn,usenatbib]{mnras}

\usepackage{newtxtext,newtxmath}

\usepackage[T1]{fontenc}

\DeclareRobustCommand{\VAN}[3]{#2}
\let\VANthebibliography\thebibliography
\def\thebibliography{\DeclareRobustCommand{\VAN}[3]{##3}\VANthebibliography}

\usepackage{graphicx}	
\usepackage{amsmath}	

\title[Data cube treatment on kinematic analyses]{Evaluating the efficacy of a data cube treatment procedure for kinematic analyses: application to NGC 3115 and NGC 4699}

\author[R. B. Menezes et al.]{
R. B. Menezes$^{1}$\thanks{\href{mailto:roberto.menezes@maua.br}{roberto.menezes@maua.br}},
L. D. B. Sonoda$^{1}$\thanks{\href{mailto:22.01716-0@maua.br}{22.01716-0@maua.br}}, 
Patr\'icia da Silva,$^{2}$\thanks{\href{mailto:p.silva2201@gmail.com}{p.silva2201@gmail.com}},
A. T. Monteiro$^{1}$,
T. V. Ricci$^{3}$,
R. G. Bravo$^{1}$,
\newauthor{D. D. V. Gueter$^{1}$ and
V. C. Parro$^{1}$}\\
$^{1}$Instituto Mau\'a de Tecnologia, Pra\c{c}a Mau\'a 1, 09580-900, S\~ao Caetano do Sul, SP, Brazil\\
$^{2}$Instituto de Astronomia, Geof\'isica e Ci\^encias Atmosf\'ericas, Departamento de Astronomia, Universidade de S\~ao Paulo, 05508-090, SP, Brazil\\
$^{3}$Universidade Federal da Fronteira Sul, Cerro Largo, 97900000, RS, Brazil
}

\date{Accepted 2025 September 05. Received 2025 August 07; in original form 2025 March 19}

\pubyear{\the\year{}}

\begin{document}
\label{firstpage}
\pagerange{\pageref{firstpage}--\pageref{lastpage}}
\maketitle

\begin{abstract}

Data cubes have been increasingly used in astronomy. These data sets, however, are usually affected by instrumental effects and high-frequency noise. In this work, we evaluate the efficacy of a data cube treatment methodology, previously proposed by our research group, for analyses focused on the stellar and gas kinematics. To do that, we used data cubes of the central regions of the galaxies NGC 3115 and NGC 4699, obtained with the Integral Field Unit of the Gemini Multi-Object Spectrograph. For each galaxy, we analysed three data cubes: non-treated, filtered (with the Butterworth spatial filtering) and filtered and deconvolved (with the Richardson-Lucy deconvolution). For each data cube, we performed a dynamical modelling, using Jeans Anisotropic Models, to obtain, among other parameters, the masses of the central supermassive black holes. Both for NGC 3115 and NGC 4699, the values of the parameters provided by the dynamical modelling from the non-treated, filtered and filtered and deconvolved data cubes were compatible, at the 1-$\sigma$ level. However, the use of the Butterworth spatial filtering decreased the uncertainty of the parameters. The additional use of the Richardson-Lucy deconvolution decreased even more the uncertainty of the parameters. The complete data treatment procedure resulted in decreases of 41\% and 45\% in the uncertainties of the supermassive black hole masses in NGC 3115 and NGC 4699, respectively. These results indicate that our treatment procedure not only does not compromise analyses of data cubes focused on the stellar or gas kinematics, but actually improves the quality of the results.

\end{abstract}

\begin{keywords}
techniques: imaging spectroscopy -- galaxies: individual: NGC 3115 -- galaxies: individual: NGC 4699 -- galaxies: nuclei
\end{keywords}



\section{Introduction}\label{sec1}

\begin{small}
\begin{table*}
\centering
\caption{Details of the observations of the four galaxies analysed in this work. \label{tbl1}}
\begin{tabular}{ccccccc}
\hline
Galaxy & Programme ID (PI) & Number & Exposure & Grating & FWHM$_{5500}$ (arcsec) of & FWHM$_{5500}$ (arcsec) of \\
       &          & of exposures & time (s)&  & the non-treated data cube & the treated data cube
\\ \hline
NGC 3115 & GS-2013A-Q-52 (T. V. Ricci) & 1 & 1800 & B600-G5323 & 0.66 & 0.52 \\
NGC 4699 & GS-2013A-Q-52 (T. V. Ricci) & 1 & 1800 & B600-G5323 & 0.43 & 0.34 \\

\hline
\end{tabular}
\end{table*}
\end{small}

Astronomical data have increased both in volume and complexity in the last decades. This is especially true if we consider analysing data cubes of large samples of astronomical objects. Data cubes are essentially data sets with two spatial dimensions and one spectral dimension, which provide images of the same object at different wavelengths and spectra of different spatial regions of an object. With the advent of instruments such as Integral Field Units (IFUs) and Fabry-Perot spectrographs, the use of data cubes has become more and more frequent in astronomy. Considering the amount of information contained in data cubes, adequate analyses of these data sets can provide relevant results for different sub-areas in astronomy. On the other hand, data cubes are usually affected by instrumental effects and high-frequency noise, which can compromise the quality of the data analysis. So, in order to better benefit from this kind of data, it may be necessary to use appropriate treatment procedures to enhance the quality and the amount of information one can get.

\begin{figure*}
    \centering
    \includegraphics[scale=0.72]{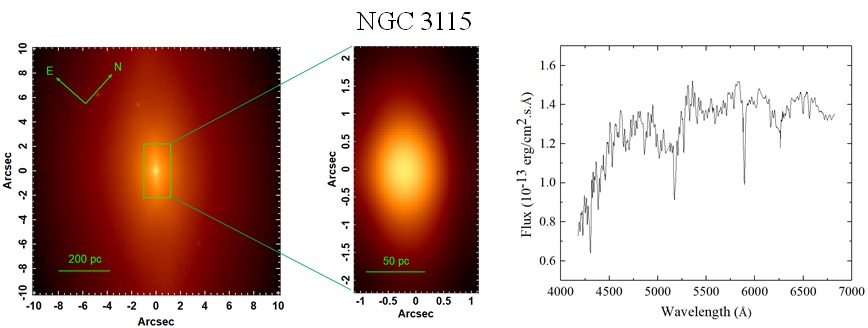}
    \includegraphics[scale=0.72]{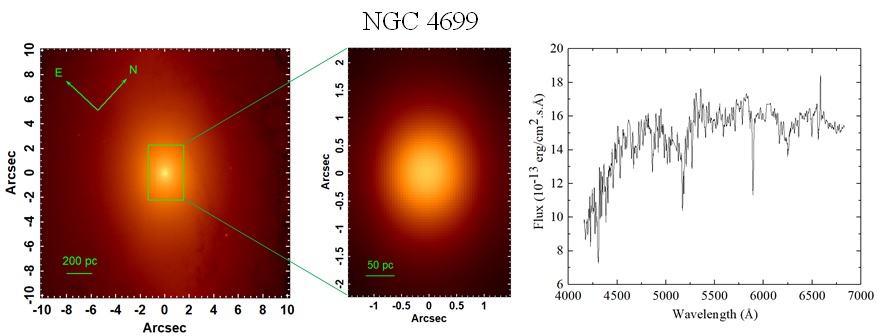}
    \caption{Left: HST image, in the F814W filter, of each galaxy analysed in this work, with the GMOS FOV marked as a rectangle. Centre: image of the treated data cube of each galaxy, collapsed along the spectral axis. Right: average spectrum of the treated data cube of each galaxy.}
    \label{fig1}
\end{figure*}

In \citet{men14a} and \citet{men15a,men19}, we presented a treatment procedure to be applied to data cubes obtained with the following instruments: Near-Infrared Integral Field Spectrograph, at the Gemini-North telescope; Spectrograph for Integral Field Observations in the Near Infrared (SINFONI), at the Very Large Telescope; Gemini Multi-Object Spectrograph (GMOS), at the Gemini-North and Gemini-South telescopes. In principle, the methodology could actually be applied to data cubes obtained with any instrument. The treatment procedure involves the following steps: correction of the differential atmospheric refraction \citep{fil82}, which is essential, as this effect may result in a displacement of the spatial structures along the spectral axis of a data cube; Butterworth spatial filtering \citep{gon02}, to remove high-frequency noise from the images of a data cube; instrumental fingerprint removal, which involves applying the Principal Component Analysis Tomography \citep{ste09} and may be necessary in certain situations, as these fingerprints can compromise different analyses to be applied to data cubes; Richardson-Lucy deconvolution \citep{ric72,luc74}, to improve the spatial resolution of the images of a data cube.

The improvements in the quality of images of data cubes provided by our treatment methodology were presented in detail in \citet{men14a,men15a,men19}. Such improvements are particularly relevant for analyses focused on the morphology of different emitting spatial structures in a data cube. On the other hand, the effects of our treatment methodology on the stellar or gas kinematics have not been analysed in detail so far. Therefore, in this work, in order to evaluate the impact of the treatment procedure on analyses involving (stellar or gas) kinematics in data cubes, we estimate, using stellar dynamical modelling, the masses of the central supermassive black holes (SMBHs) in the optical data cubes, before and after the treatment procedure, of the nuclear regions of two nearby galaxies: NGC 3115 and NGC 4699. If the treatment methodology is effective and does not compromise the data, the SMBH mass values obtained before and after the data treatment must be compatible and the uncertainties obtained from the treated data should be lower than those corresponding to the non-treated data.

The determination of masses of SMBHs is certainly a very important research topic in the current astrophysics. It is widely accepted that SMBHs, with masses between $10^6$ M$_{\sun}$ and $10^{10}$ M$_{\sun}$, are present in the nuclei of all massive galaxies \citep{kor95,ric98}. The masses of these SMBHs show correlations with certain properties of the host galaxies, such as the stellar velocity dispersion, which results in the so-called $M_{BH}$ - $\sigma$ relation \citep{fer00,geb00,gul09}. A detailed modelling of the stellar or gas kinematics around a SMBH is usually required, in order to estimate its mass. The procedure consists of trying to reproduce the observed kinematics, assuming different values for a number of parameters, including the SMBH mass.

NGC 3115 is an S0 galaxy at a distance of 9.68 Mpc \citep{ton01}. A number of previous works analysed not only the nuclear region of this object, but also its outskirts (e.g. \citealt{nor06,shc14,gue16,alm18,buz21}). A compact radio emission, with a luminosity of $3.1 \times 10^{35}$ erg s$^{-1}$, coincident with the optical nucleus, was first detected by \citet{wro12}, suggesting the presence of a low luminosity Active Galactic Nucleus (AGN) in this galaxy. This is consistent with an X-ray candidate nucleus detected by \citet{won11}. \citet{men14b} reported the detection of a broad H$\alpha$ emission line, with a luminosity of $(4.2 \pm 0.4) \times 10^{37}$ erg s$^{-1}$, at a projected distance of $\sim0.29 \pm 0.05$ arcsec (corresponding to $\sim14.3 \pm 2.5$ pc) from the centre of the stellar bulge, which suggests the presence of an off-centred AGN in NGC 3115. This result is consistent with two possible scenarios: the observed off-centred AGN is part of a SMBH binary system, with a second SMBH at the centre of the stellar bulge; the off-centred AGN is the result of the coalescence of two SMBHs (resulting probably from a merger) and was ``kicked'' from the centre of the stellar bulge. In a latter study, \citet{jon19} concluded that the radio source in this galaxy coincides with the stellar bulge centre, which is an argument against the scenario involving the ``kicked'' merger remnant AGN, but not against the scenario with a binary SMBH system. NGC 3115 is considered the nearest galaxy with a central SMBH with a mass of $\sim10^9$ M$_{\sun}$. Such a mass was estimated in a number of previous studies, based on the modelling of the stellar kinematics (e.g. \citealt{kor92,kor96,ems99}). \citet{mag98} obtained a somewhat lower value of $4.06^{+0.19}_{-0.22} \times 10^8$ $M_{\sun}$ for the SMBH mass.

NGC 4699 is classified as an SAB(rs)b galaxy, at a distance of 19.5 Mpc \citep{tul16}. There is not much information in the literature about the nuclear region of this object. \citet{bow93} analysed the stellar rotation velocity and the stellar velocity dispersion along the major and minor axes of NGC 4699 and established an upper limit of $3 \times 10^9$ M$_{\sun}$ for the mass of the central SMBH. Modelling of the stellar kinematics in this object has also been performed using high spatial resolution adaptive optics (AO) SINFONI data \citep{erw15,sag16}, which provided a SMBH mass of $(1.76 \pm 0.23) \times 10^8$ M$_{\sun}$.

NGC 3115 and NGC 4699 are part of the Deep IFS View of Nuclei of Galaxies (DIVING$^{3D}$) survey \citep{ste22,men22,ric23}, which is being conducted by our research group and has the goal of observing, using seeing-limited optical 3D spectroscopy, the central regions of all galaxies in the Southern hemisphere with $B$ < 12.0 and |$b$| > 15$\degr$. The complete sample has a total of 170 objects. We chose these two objects to be included in this work because both show maps of the stellar radial velocities in their nuclear regions without apparent deviations from a pure rotational pattern. The same applies to the maps of the stellar velocity dispersion in the central regions of these objects, showing a central maximum without many irregularities, which would be difficult to reproduce with the dynamical modelling we use.

As mentioned above, the main goal of this paper is to evaluate the efficiency of our treatment methodology for analyses of data cubes focused on the stellar and gas kinematics of the objects. To do that, we estimated the masses of the SMBHs, using the surrounding stellar kinematics, in the data cubes of the central regions of NGC 3115 and NGC 4699, before and after the use of our treatment procedure. This paper is organized as follows. In Section 2, we present the details of the observations of the two analysed galaxies and also the steps of the treatment procedure. In Section 3, we describe the entire data analysis and also present the results. In Section 4, we discuss and compare our results with previous works. We present our conclusions in Section 5.

\section{Observations, data reduction and data treatment}\label{sec2}

The observations of the central regions of NGC 3115 and NGC 4699 were made with the IFU of the GMOS, at the Gemini-South telescope. For each galaxy, one 30 min exposure was taken, in the one-slit mode, with the B600-G5323 grating, which resulted in a data cube with a field of view (FOV) of 5$\arcsec$ $\times$ 3.5$\arcsec$, a spectral coverage of 4250 \AA - 7000 \AA~and a spectral resolution of 1.8 \AA, based on the sky emission line [O \textsc{i}]$\lambda$5577. The details of all observations are shown in Table~\ref{tbl1}.

The data reduction was performed in \textsc{iraf} (\textsc{image reduction and analysis facility}) environment, with the Gemini package, and included the following basic steps: trimming and bias subtraction; cosmic ray removal with the \textsc{l. a. cosmic} routine \citep{van01}, which was essential, as only one exposure was taken for each object and, as a consequence, it was not possible to calculate medians of the data cubes to remove cosmic rays; bad pixel correction, based on a bad pixel mask obtained using GCAL-flat images; extraction of the spectra; correction for pixel to pixel gain variations, using response curves also obtained from GCAL-flat images; wavelength calibration, applied based on a wavelength solution provided by CuAr lamp images; subtraction of the sky emission, using the average spectrum obtained from the GMOS sky FOV (at a distance of 1 arcmin of the science FOV); flux calibration, using data of the observed spectrophotometric standard stars; construction of the data cubes, with spatial pixels (spaxels) of 0.05$\arcsec$.

As explained in Section~\ref{sec1}, the reduced data cubes were treated with a procedure including correction of the differential atmospheric refraction, Butterworth spatial filtering, instrument fingerprint removal and Richardson-Lucy deconvolution. The Butterworth spatial filtering is applied to each image of the data cubes, to remove high spatial-frequency noise. In the case of GMOS/IFU data cubes, such a noise is probably related to the lenslet/fibre mechanism of the instrument, although any spatial re-sampling applied to the data cubes may also introduce noise. As explained in \citet{men19}, the Butterworth spatial filtering applied to GMOS/IFU data cubes using, as a filter, the product of two identical circular filters, with order $n$ = 2 and cut-off frequencies between 0.16 and 0.25 Ny, removes most of the noise, without introducing artefacts. For data cubes in general (obtained with any instrument), higher order filters usually introduce concentric rings around point-like sources in the images. The adequate cut-off frequency to be used for the filtering should remove most of the noise, without affecting the point spread function (PSF) of the observation. For the data cubes of NGC 3115 and NGC 4699, the cut-off frequencies used in the filtering procedure were 0.19 Ny and 0.20 Ny, respectively. 

The results obtained with the Richardson-Lucy deconvolution applied to the images of data cubes are highly dependent on the accuracy of the PSF estimate and on the number of iterations used in the procedure \citep{men14a,men15a,men19}. A deconvolution applied with less than six iterations will not result in a significant improvement in the spatial resolution of the observation. On the other hand, a number of iterations higher than ten will amplify the low-frequency spatial noise in the images. In order to obtain a reliable estimate of the PSF in a data cube, one may use, for example, the acquisition image of the observation, possible stars in the FOV, data cubes of standard stars (if observed not long before or after the science data cube) or even the image of the broad component of permitted emission lines (when present) in AGNs, which should be point-like. Better results are obtained if the variation of the PSF with wavelength is taken into account. For the data cubes of NGC 3115 and NGC 4699, the Richardson-Lucy deconvolution was applied with six iterations, using a Gaussian PSF and assuming that the full width at half maximum (FWHM) of the PSF varies along the spectral axis of the data cube according to the following equation (which was derived from the analysis of GMOS/IFU standard stars data cubes):

\begin{equation}
FWHM \left(\lambda\right) = FWHM_{ref} \cdot \left(\frac{\lambda}{\lambda_{ref}}\right)^{\alpha},
\label{eq1}
\end{equation}\\
where FWHM$_{ref}$ is the FWHM at $\lambda_{ref}$. 

For all the objects analysed in this work, we assumed $\alpha$ = -0.3 (usually, for GMOS/IFU data cubes, the $\alpha$ values are in the range of -0.35 to -0.25) and $\lambda_{ref}$ = 6300 \AA~(which corresponds to the wavelength of the acquisiton images). The values of the FWHM$_{ref}$ were estimated from the acquisition images. The values of the FWHM at 5500 \AA~of the non-deconvolved data cubes (Table~\ref{tbl1}) were then calculated, using equation 1. The FWHM values at 5500 \AA~of the deconvolved data cubes (Table~\ref{tbl1}) were estimated as being equal to 79\% of the corresponding values of the non-deconvolved data cubes \citep{men19,men22}. For further detail about the Butterworth spatial filtering and the Richardson-Lucy deconvolution, see \citet{men14a,men15a,men19}.

Fig.~\ref{fig1} shows \textit{Hubble Space Telescope} (\textit{HST}) images of NGC 3115 (Instrument: WFPC2; PI last name: Faber; Proposal ID: 5512; Exptime: 80.0 s) and NGC 4699 (Instrument: WFC3; PI last name: Erwin; Proposal ID: 15133; Exptime: 500.0 s), in the F814W filter, together with images and average spectra of the treated GMOS/IFU data cubes of these two objects.

\begin{figure*}
    \centering
    \includegraphics[scale=0.53]{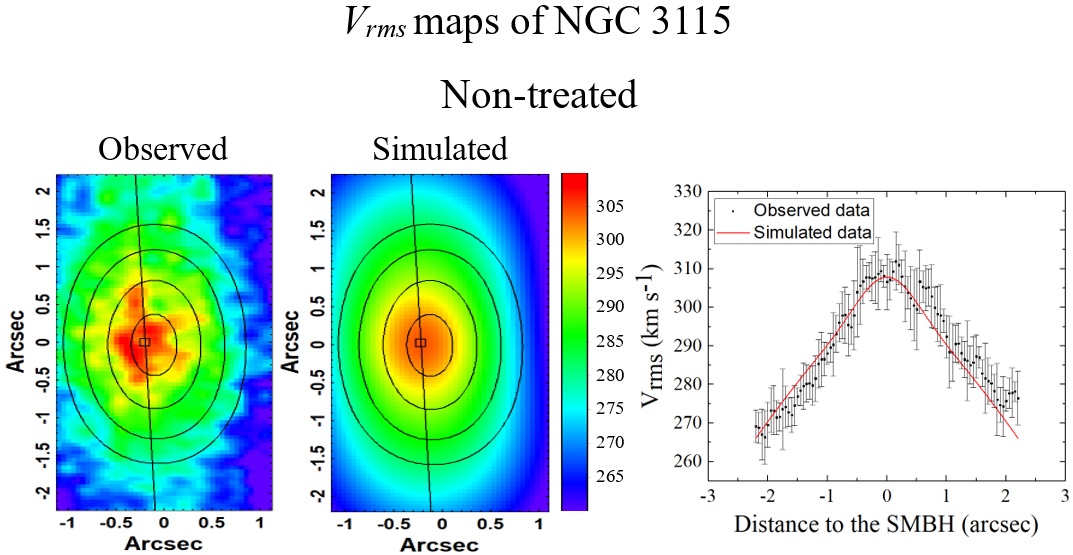}
    \includegraphics[scale=0.53]{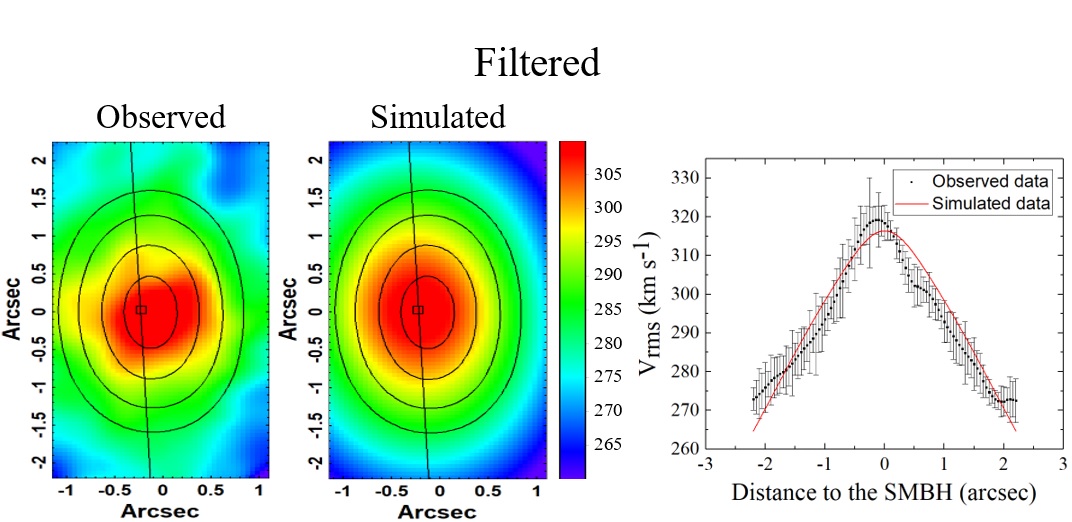}
    \includegraphics[scale=0.53]{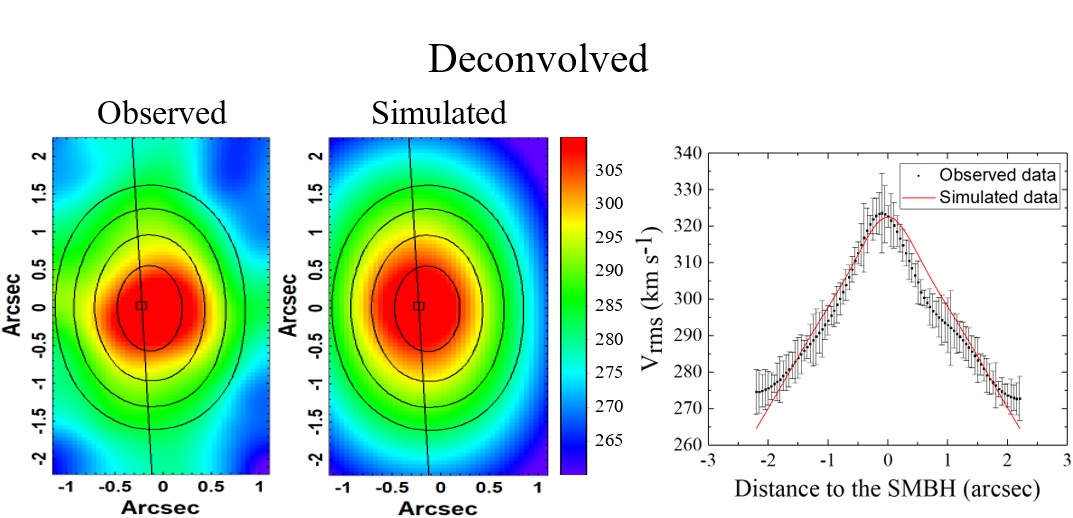}
    \caption{$V_{rms}$ maps (left) of the non-treated, filtered and deconvolved data cubes of NGC 3115, obtained with the pPXF technique, together with the simulated $V_{rms}$ maps (centre) corresponding to the best obtained models and the curves (right) extracted along the kinematic axis of the maps (shown as a black line). The position of the central SMBH (coincident with the stellar nucleus) is marked with a square. The elliptical black curves correspond to the isocontours of the data cubes, integrated along the spectral axis. All $V_{rms}$ values are in km s$^{-1}$.
    }
    \label{fig2}
\end{figure*}

\begin{figure*}
    \centering
    \includegraphics[scale=0.53]{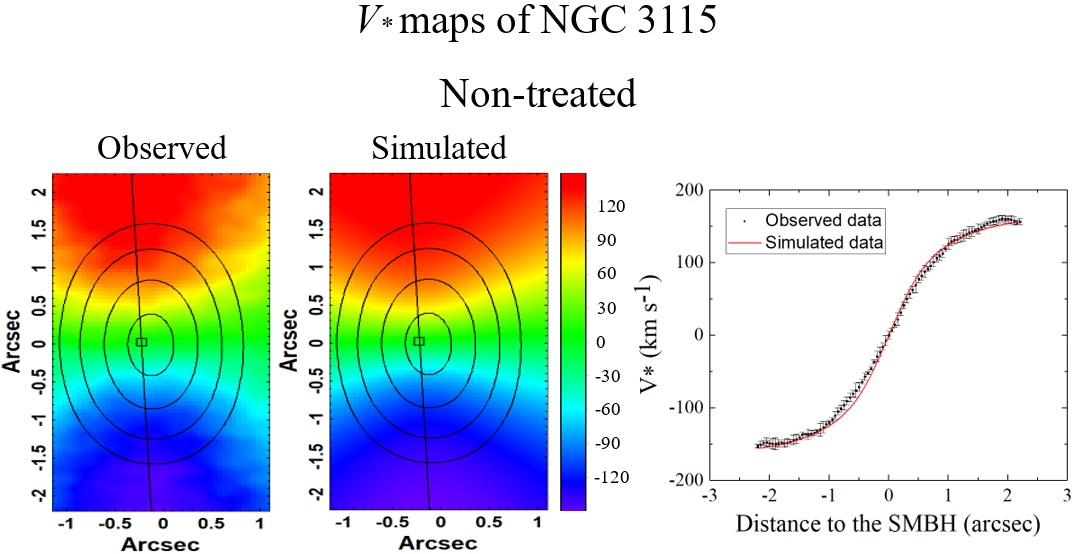}
    \includegraphics[scale=0.53]{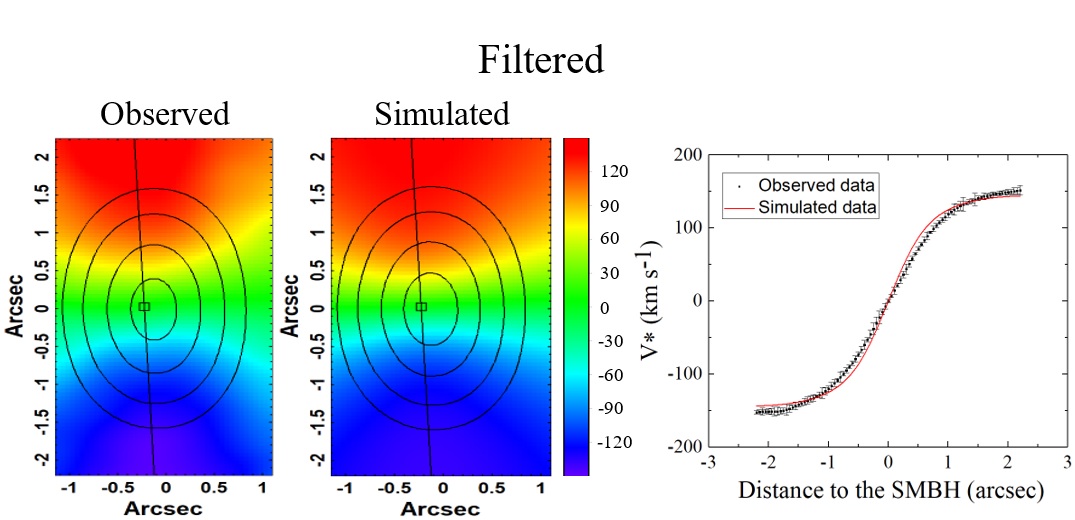}
    \includegraphics[scale=0.53]{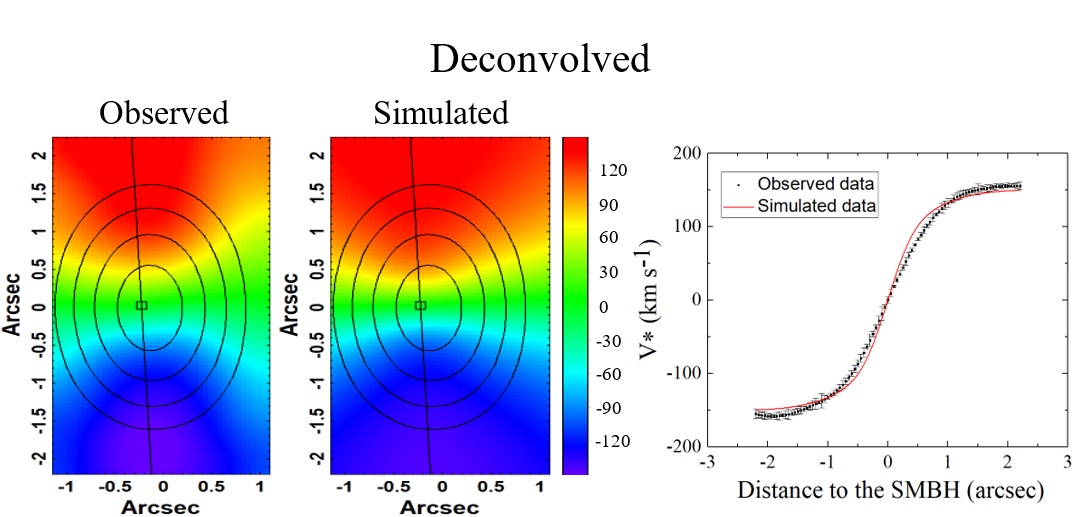}
    \caption{The same as in Fig.~\ref{fig2}, but for the $V_*$ maps of NGC 3115.}
    \label{fig3}
\end{figure*}

\begin{figure*}
    \centering
    \includegraphics[scale=0.53]{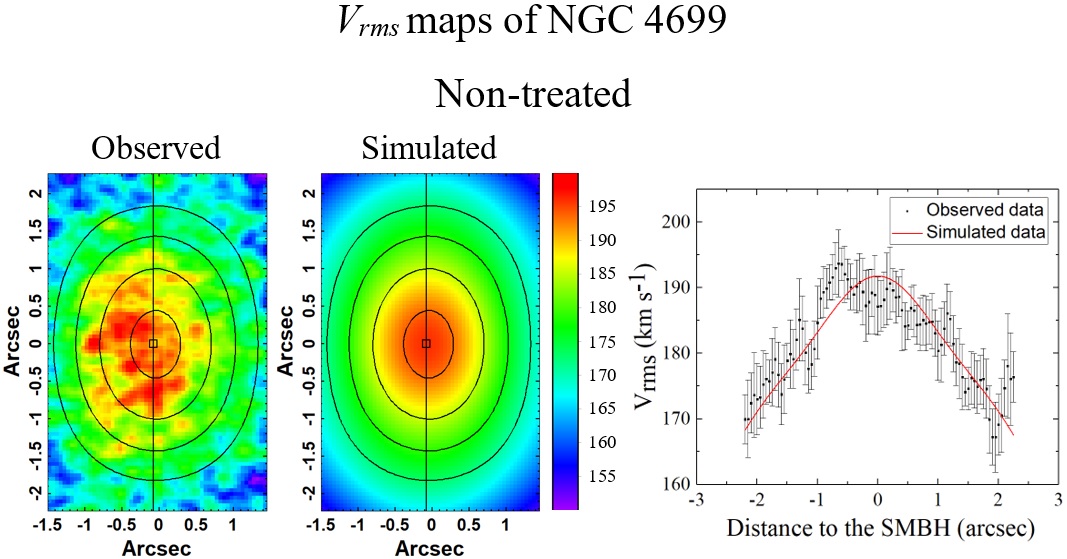}
    \includegraphics[scale=0.53]{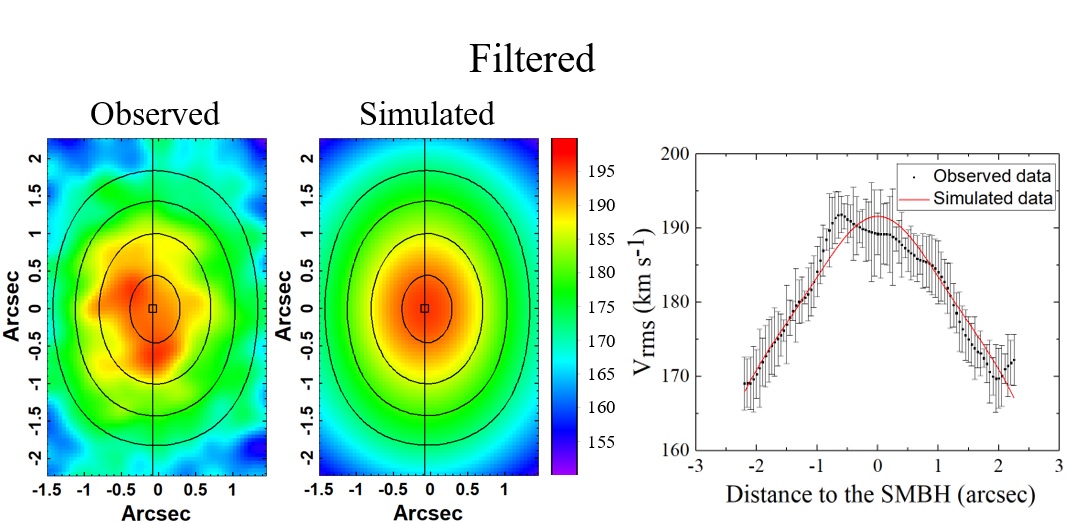}
    \includegraphics[scale=0.53]{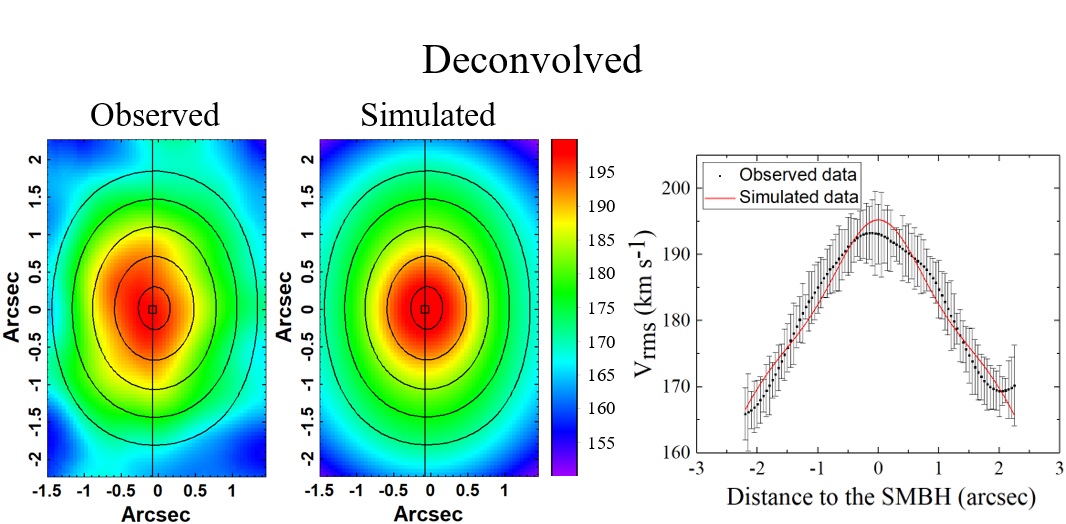}
    \caption{The same as in Fig.~\ref{fig2}, but for the $V_{rms}$ maps of NGC 4699.}
    \label{fig4}
\end{figure*}

\begin{figure*}
    \centering
    \includegraphics[scale=0.53]{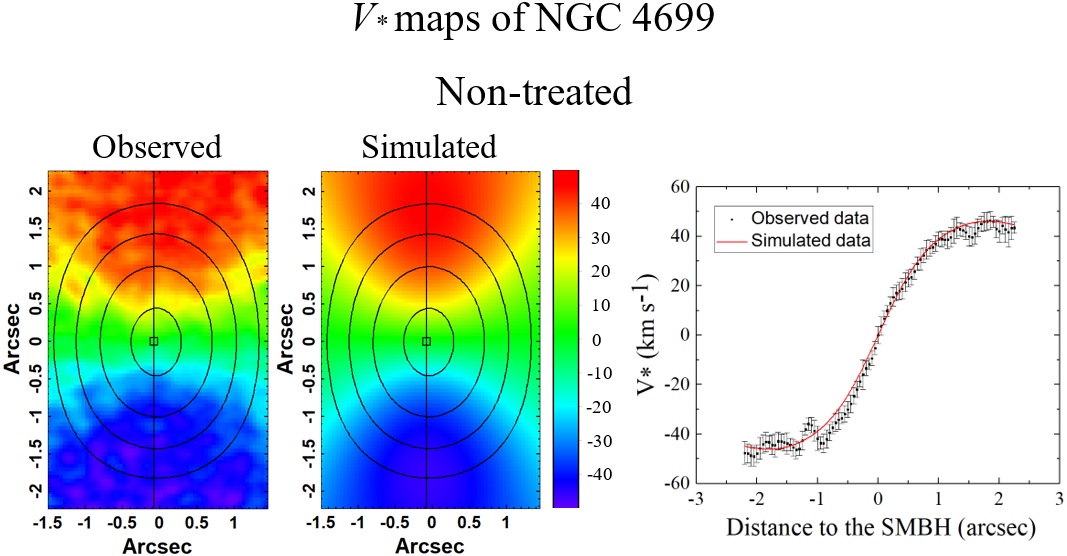}
    \includegraphics[scale=0.53]{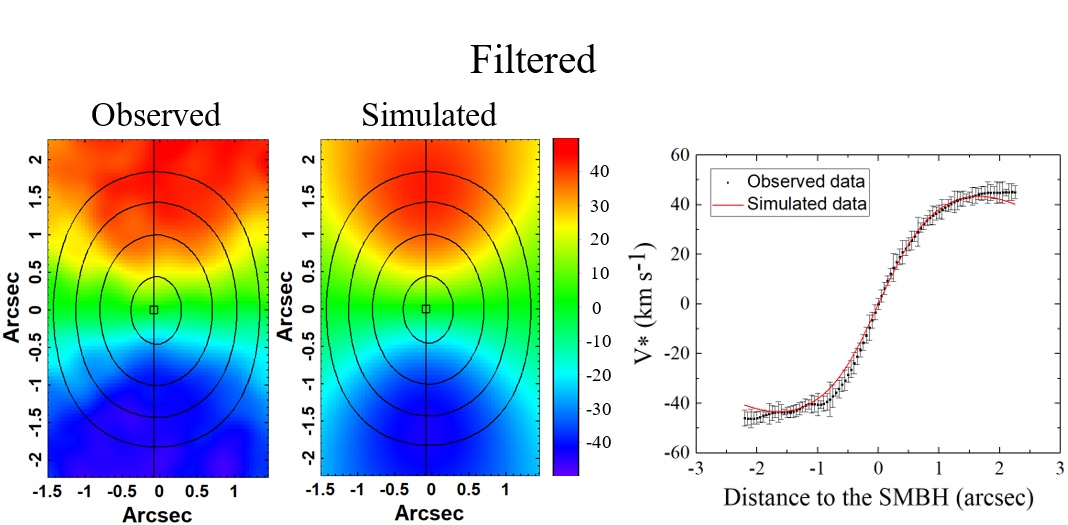}
    \includegraphics[scale=0.53]{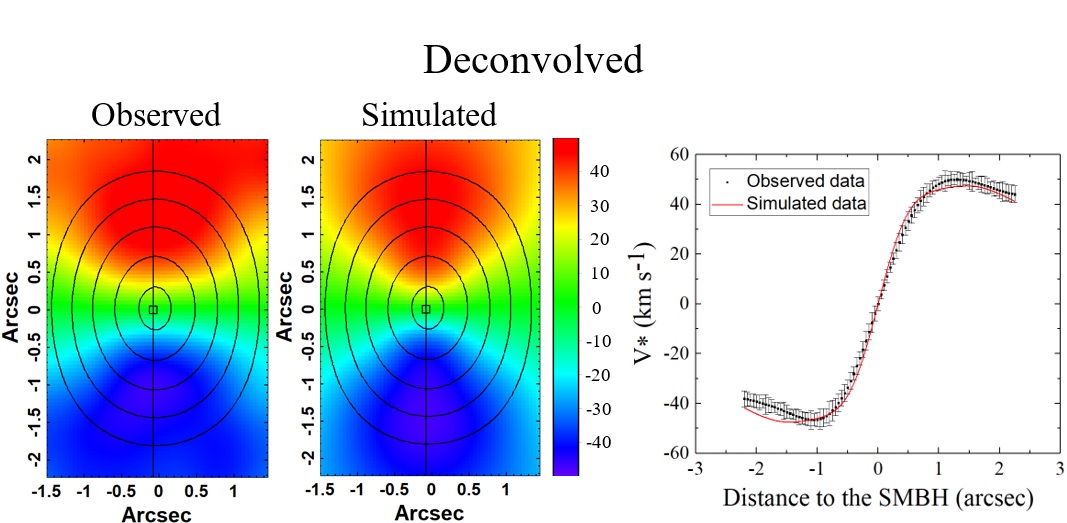}
    \caption{The same as in Fig.~\ref{fig2}, but for the $V_*$ maps of NGC 4699.}
    \label{fig5}
\end{figure*}

\begin{figure*}
    \centering
    \includegraphics[scale=0.48]{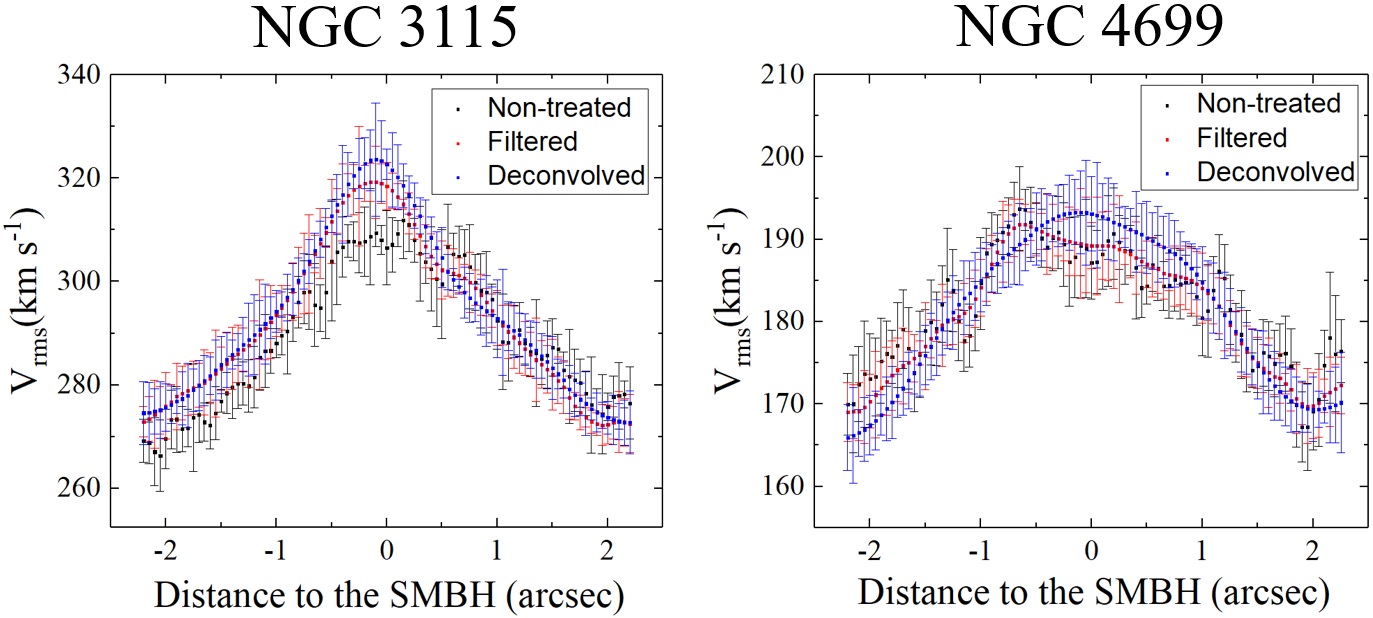}
    \caption{Curves of the $V_{rms}$ maps shown in Figs.~\ref{fig2} and ~\ref{fig4}, extracted along the corresponding kinematic axes.}
    \label{fig6}
\end{figure*}

\begin{figure*}
    \centering
    \includegraphics[scale=0.48]{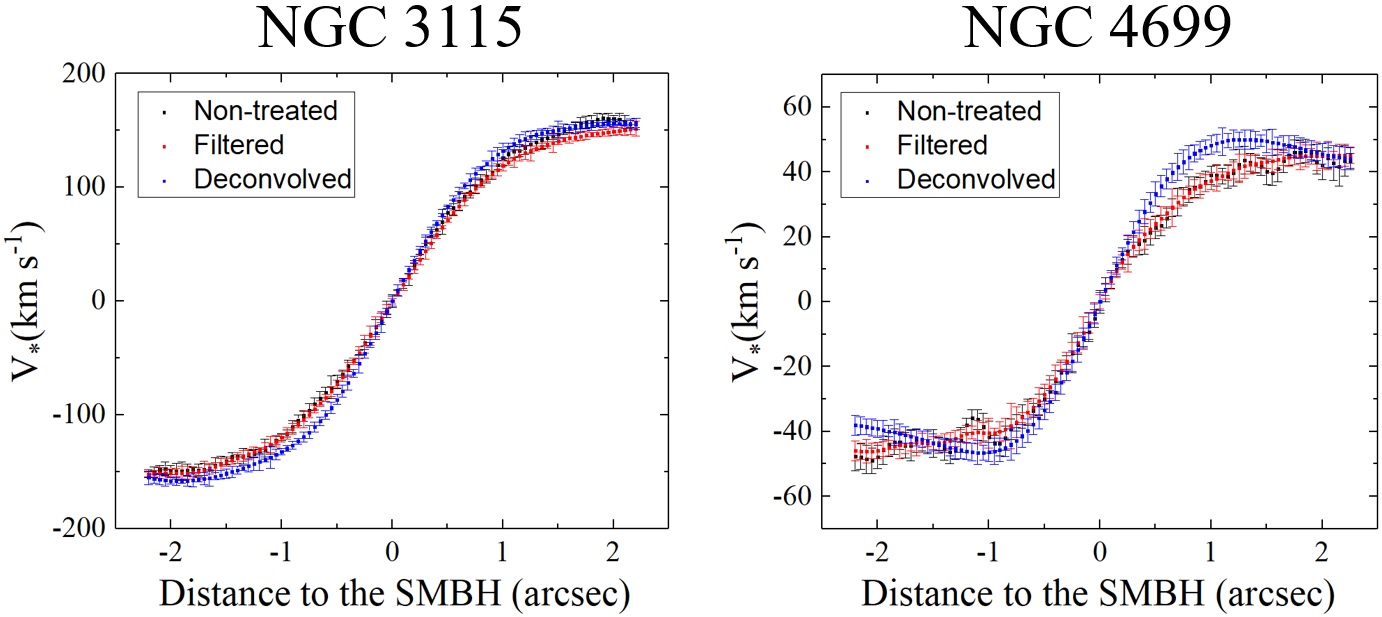}
    \caption{Curves of the $V_*$ maps shown in Figs.~\ref{fig3} and ~\ref{fig5}, extracted along the corresponding kinematic axes.}
    \label{fig7}
\end{figure*}

Using the SMBH mass of NGC 3115 obtained by \citet{mag98} and a central stellar velocity dispersion of $260 \pm 3$ km s$^{-1}$ (from Hyperleda\footnote{http://leda.univ-lyon1.fr/} -- \citealt{hyperleda}), we obtain, considering the distance of this galaxy, a diameter for the sphere of influence of the central SMBH of $\sim$ 1.11\arcsec. Since this value is considerably higher than the FWHM of the PSF at 5500\AA~of the non-treated and treated data cubes of this galaxy, we conclude that this sphere of influence can easily be spatially resolved in this observation. On the other hand, for NGC 4699, using the SMBH mass of \citet{sag16} and \citet{erw15} and a central stellar velocity dispersion of $192 \pm 9$ km s$^{-1}$, the diameter of the sphere of influence of the central SMBH is 0.44\arcsec, which is almost equal to the FWHM of the PSF at 5500\AA~of the non-treated data cube, but is higher than the FWHM of the PSF of the treated data cube. Therefore, we conclude that the sphere of influence of the central SMBH in NGC 4699 is at the limit of the spatial resolution of the non-treated data cube, but is easier to be spatially resolved in the treated data cube. One of the reasons to choose the data cube of this galaxy for the analysis in this work is exactly to have an example at the limit of the required spatial resolution to perform the dynamical modelling.

\begin{figure*}
    \centering
    \includegraphics[scale=0.93]{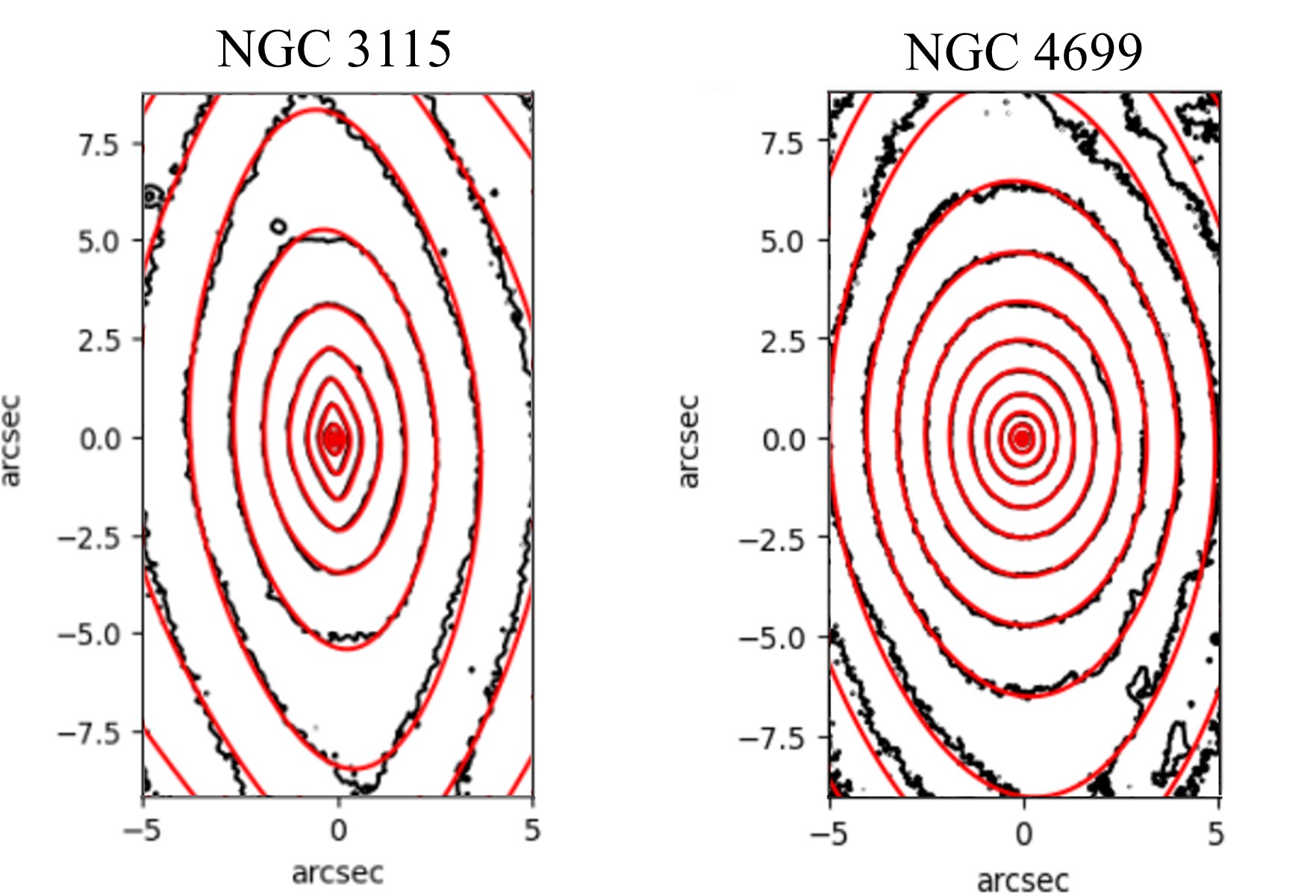}
    \caption{Fits of the \textit{HST} images of NGC 3115 and NGC 4699 provided by the MGE technique. The black lines correspond to the isophotes of the \textit{HST} images and the red lines represent the MGE fits.}
    \label{fig8}
\end{figure*}

The data analysis described in the next section was performed, for each galaxy, using three data cubes: the non-treated data cube, obtained after the data reduction and only with the correction of the differential atmospheric refraction applied (as the analysis of a data cube without this preliminary correction would not provide relevant results for this work); the filtered data cube, obtained after the Butterworth spatial filtering was applied; the deconvolved data cube, obtained after both the Butterworth spatial filtering and the Richardson-Lucy deconvolution were applied.

\section{Data analysis}\label{sec3}

\subsection{Determination of the stellar kinematic parameters}\label{sec31}

The first step of the data analysis involved the determination of the values of the stellar radial velocity ($V_*$) and of the stellar velocity dispersion ($\sigma_*$) for the analysed data cubes. This was achieved with the Penalized Pixel Fitting (pPXF) procedure \citep{cap17}, which fits a stellar spectrum with a combination of template spectra from a base, convolved with a Gauss-Hermite expansion. We opted to use a base of stellar population spectra obtained from the Medium-resolution Isaac Newton Telescope Library of Empirical Spectra (MILES; \citealt{san06}). The base was assembled with a Kroupa universal initial mass function, with a slope of 1.3, and the isochrones of the Bag of Stellar Tracks and Isochrones (BaSTI; \citealt{hid18}). The resulting base has stellar populations with ages between $3 \times 10^7$ yr and $1.3 \times 10^{10}$ yr and metallicities ([M/H]) between -0.66 to +0.40. This base was chosen mainly because the fits provided by the pPXF technique reproduced, with excellent precision, the stellar continuum and the profiles of the absorption lines in the data cubes of NGC 3115 and NGC 4699, without using additive or multiplicative Legendre polynomials (which can be included in the pPXF routine developed by \citealt{cap17}). That certainly indicates that the age and metallicity ranges covered by the base are appropriate to reproduce the spectral features of the stellar populations in the nuclei of NGC 3115 and NGC 4699.

Before pPXF was applied, the spectra of the data cubes were corrected for the redshift, using z = 0.002287 for NGC 3115 and z = 0.004650 for NGC 4699 (both values from the NASA Extragalactic Database; NED\footnote{The NASA/IPAC Extragalactic Database (NED) is funded by the National Aeronautics and Space Administration and operated by the California Institute of Technology.}), and re-sampled, so that each spectral pixel corresponded to a wavelength interval of $\Delta\lambda$ = 1\AA. The pPXF technique provided not only the values of $V_*$ and $\sigma_*$, but also of the Gauss-Hermite coefficients $h_3$ and $h_4$. We opted to include $h_3$ and $h_4$ in the procedure (in addition to $V_*$ and $\sigma_*$) because the resulting fits reproduced with more precision the profiles of the stellar absorption lines. Such Gauss-Hermite coefficients, however, are not discussed in the analysis, due to the purposes of this work. The synthetic stellar spectrum corresponding to the fit obtained by pPXF can be used, in certain analyses, for the starlight subtraction. Since pPXF was applied to the spectrum corresponding to each spaxel of the data cubes, the results were $V_*$ and $\sigma_*$ maps for each analysed data cube. All emission lines in the data cubes were masked before pPXF was applied.

For the non-treated data cube of NGC 3115, the signal-to-noise (S/N) ratio of the stellar continuum was $\sim90$ in the central regions of the FOV and $\sim60$ in areas farther from the nucleus. For the filtered and deconvolved data cubes of the same object, the S/N ratio in the nuclear region and in the outskirts of the FOV was $\sim115$ and $\sim100$, respectively. In the case of NGC 4699, the stellar continuum S/N ratio of the non-treated data cube was $\sim80$ in the central regions and $\sim50$ in the outskirts of the FOV. For the filtered and deconvolved data cubes of this galaxy, the S/N ratio was $\sim100$ and $\sim80$ in the nuclear region and in areas farther from the nucleus, respectively. All these high S/N ratio values allowed a robust fitting of all the kinematic parameters analysed in this work. 

The uncertainties of the $V_*$ and $\sigma_*$ values provided by pPXF were estimated using a Monte Carlo procedure. For each fitted spectrum, first of all, the synthetic stellar spectrum corresponding to the fit performed by pPXF was subtracted from the original data, resulting in a spectrum containing only noise and emission lines. Then, a histogram of the values (representing noise) in a spectral region without emission lines was constructed. A Gaussian function was fitted to the histogram. After that, random distributions of noise, following the Gaussian distribution fitted to the histogram, were obtained and added to the synthetic spectral fit provided by pPXF, resulting in ``noisy'' spectra. Finally, the pPXF was applied to each of the ``noisy'' spectra and the uncertainties of $V_*$ and $\sigma_*$ were taken as the standard deviations of the values obtained with each of the pPXF fits.

Using the relation $V_{rms} = \sqrt{V_*^2 + \sigma_*^2}$, we also obtained $V_{rms}$ maps for each analysed data cube. The uncertainties of $V_{rms}$ were determined using simple propagation of uncertainties of $\sigma_*$ and $V_*$. Figs.~\ref{fig2} and~\ref{fig3} show the $V_{rms}$ and $V_*$ maps, respectively, for the non-treated, filtered and deconvolved data cubes of NGC 3115. Figs.~\ref{fig4} and~\ref{fig5} show the same, but for the non-treated, filtered and deconvolved data cubes of NGC 4699. It is important to emphasize that, from now on, the term ``deconvolved data cube'' refers to the data cube obtained after both the Butterworth spatial filtering and the Richardson-Lucy deconvolution were applied. Due to possible uncertainties in the values of the redshifts used for the corrections of the data cubes, for each $V_*$ map, we subtracted from the entire map the $V_*$ value at the position of the stellar nucleus. In other words, the $V_*$ maps in Figs.~\ref{fig3} and~\ref{fig5} show actually the stellar radial velocities relative to the stellar nucleus. It is easy to see that the morphologies of all maps are entirely consistent; however, the maps from non-treated data cubes are ``noisier'', showing high spatial-frequency structures not seen in the maps from the filtered data cubes. This was actually expected. The Butterworth spatial filtering removes high spatial-frequency components from the images of the data cubes, which also results in a decrease of the spectral noise. Such a decrease avoids possible instabilities during the pPXF spectral fitting, resulting in less ``noisy'' kinematic maps. The morphologies of the maps obtained from the deconvolved data cubes are also consistent with those of the other maps, but the differences between the maps from the filtered and deconvolved data cubes are more difficult to detect, just by looking at the maps.

Figs.~\ref{fig6} and~\ref{fig7} show the $V_{rms}$ and $V_*$ curves extracted from the kinematic axes (which correspond to the lines connecting the points with the maximum and minimum values in the $V_*$ maps) for the non-treated, filtered and deconvolved data cubes of NGC 3115 and NGC 4699. In each graph, all the values of the three curves are compatible, at the 1-$\sigma$ or 2-$\sigma$ levels. However, there are slight differences between the general profiles of the curves. Similarly to what was previously discussed, the curves of the maps from the filtered data cubes show less irregularities than those of the non-treated data cubes. The curves of the maps from the deconvolved data cubes show profiles consistent with a higher spatial resolution, in comparison to the others, which was expected, as the purpose of the Richardson-Lucy deconvolution is to improve the spatial resolution of the observations.

\subsection{Jeans Anisotropic Modelling}\label{sec32}

After the determination of the stellar kinematic parameters of the data cubes, the next step in the data analysis was the use of the Multi-Gaussian Expansion (MGE; \citealt{cap02}) fitting technique on the \textit{HST} images of NGC 3115 and NGC 4699, shown in Fig.~\ref{fig1}. This is a necessary step, in order to obtain the stellar mass profiles of the galaxies, which are required to perform a dynamical modelling. MGE essentially fits an image of a galaxy with a sum of Gaussian functions convolved with an estimate of the PSF of the observation. We opted to apply the MGE technique to \textit{HST} images taken in the F814W filter (similar to the $I$ band), in order to avoid, as much as possible, the effects of the interstellar extinction. The parameters provided by MGE, for each galaxy, were the brightness surface density in the $I$ band (in $L_{\sun,I}$ pc$^{-2}$), the width and the axial ratio of all Gaussians used to fit the image. Fig.~\ref{fig8} shows the fits of the \textit{HST} images of NGC 3115 and NGC 4699 obtained with the MGE procedure.

\begin{figure*}
    \centering
    \includegraphics[scale=0.47]{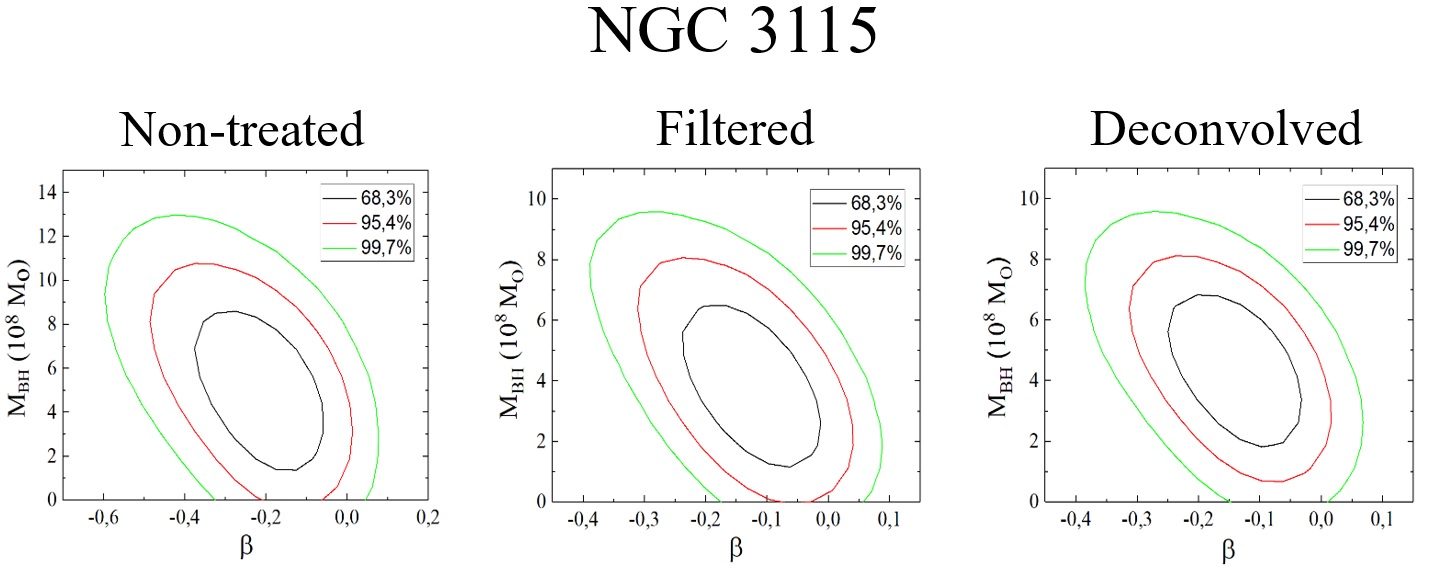}
    \includegraphics[scale=0.47]{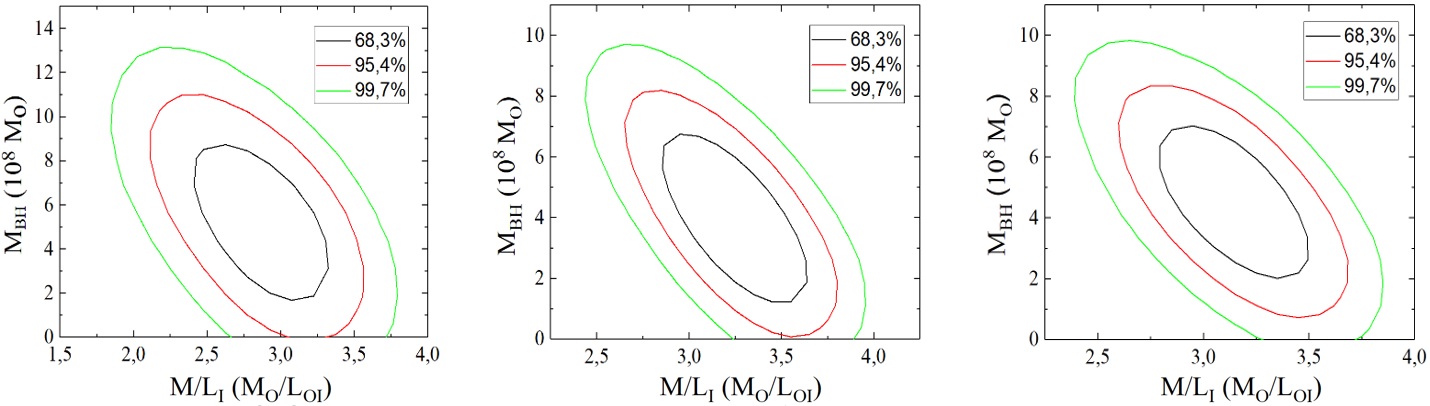}
    \includegraphics[scale=0.47]{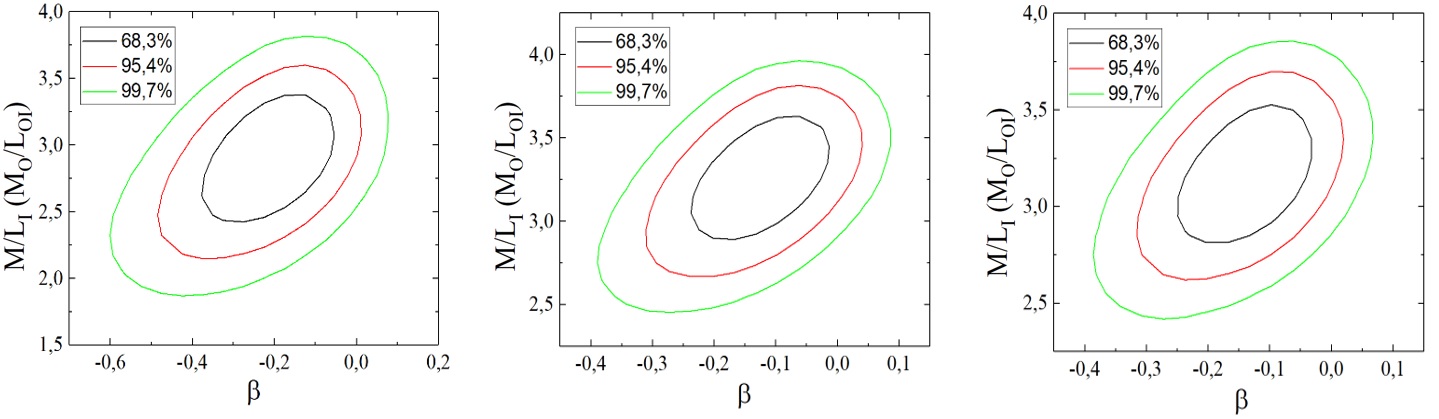}
    \caption{Posterior distributions of the dynamical modellings applied to the non-treated, filtered and deconvolved data cubes of NGC 3115. The black, red, and green contours correspond to 1-$\sigma$ (68.3\%), 2-$\sigma$ (95.4\%), and 3-$\sigma$ (99.7\%) errors, respectively.}
    \label{fig9}
\end{figure*}

\begin{figure*}
    \centering
    \includegraphics[scale=0.47]{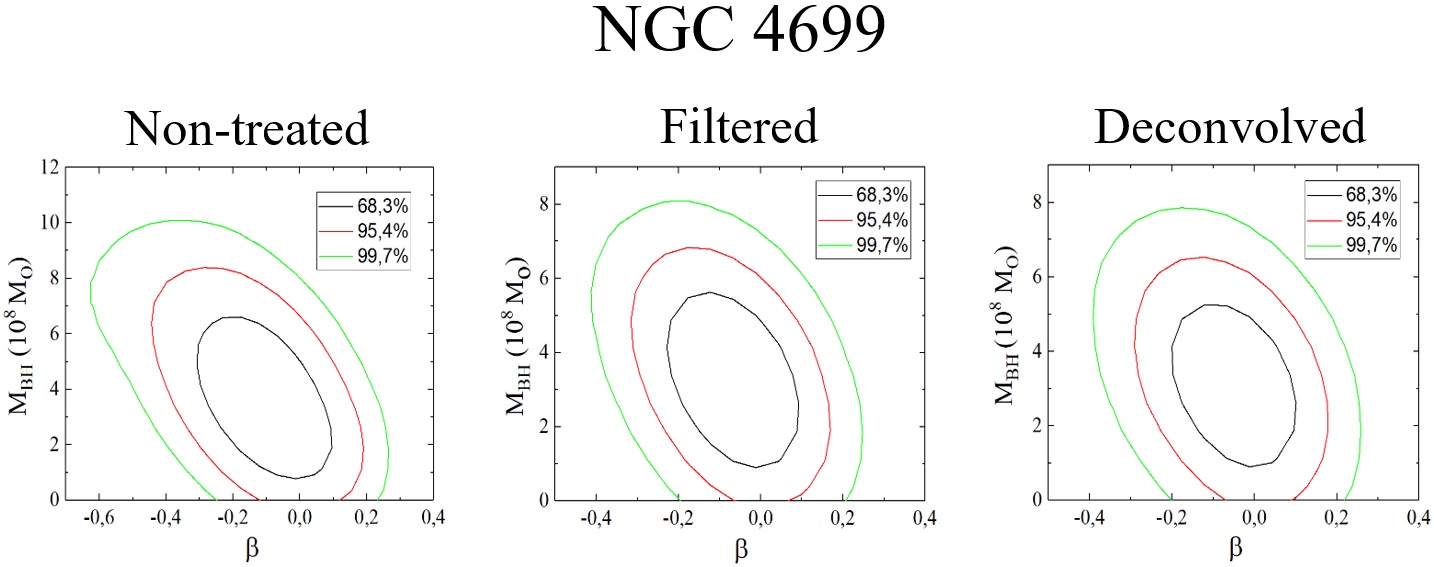}
    \includegraphics[scale=0.47]{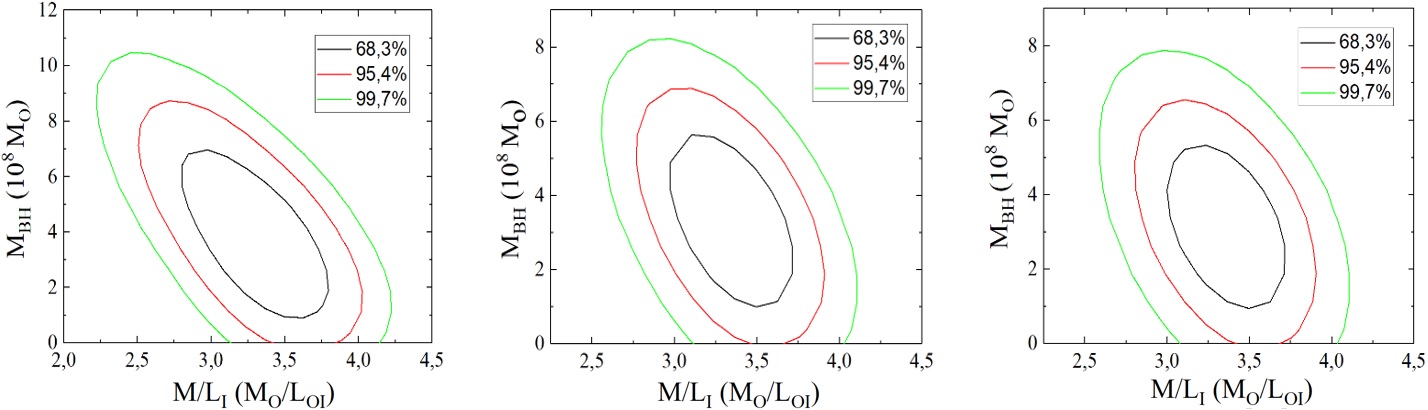}
    \includegraphics[scale=0.468]{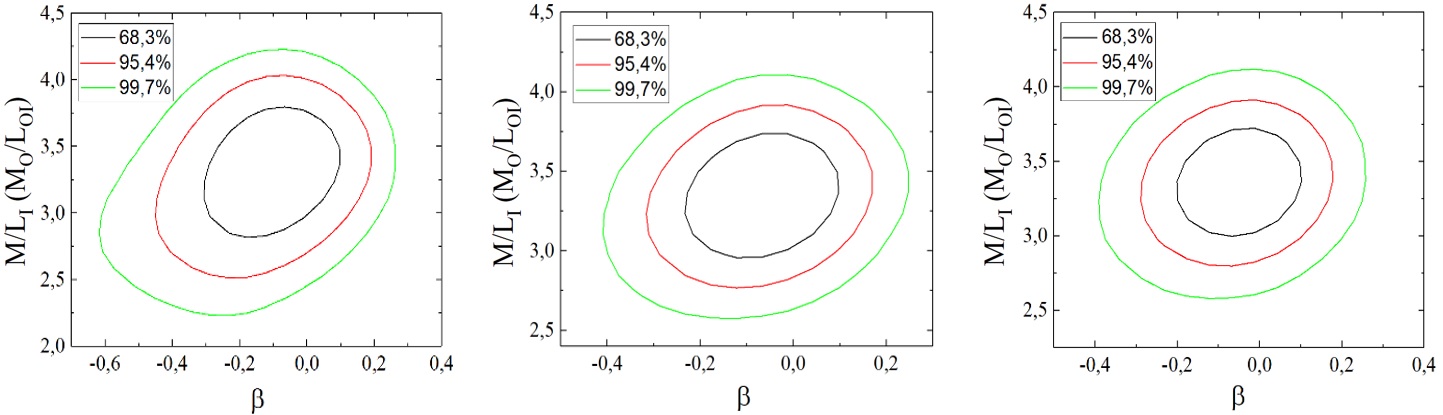}
    \caption{The same as in Fig.~\ref{fig9}, but for the non-treated, filtered and deconvolved data cubes of NGC 4699.}
    \label{fig10}
\end{figure*}

We then applied Jeans Anisotropic Modelling (\textsc{jam}) to the kinematic maps of the non-treated, filtered and deconvolved data cubes of NGC 3115 and NGC 4699, in order to establish a comparison of the parameters provided by the modelling for each of the data cubes. We used the \textsc{jam} implementation developed by \citet{cap08}, in \textsc{python} language, assuming a cylindrically-aligned velocity ellipsoid. All simulations were performed considering three free parameters: the mass of the SMBH ($M_{BH}$), the mass-to-luminosity ratio of the stellar populations in the $I$ band ($M/L_I$) and the anisotropy parameter $\beta = 1 - \sigma_z^2/\sigma_R^2$. We did not consider the inclination of the rotating stellar disc ($i$) as a free parameter, because there was only a few acceptable $i$ values for the script (due to the constraints from the MGE fit) and changing them did not alter significantly the results. We assumed $i$ = 89\degr~for all NGC 3115 simulations and $i$ = 86\degr~for all NGC 4699 simulations. Using the free parameters mentioned above, each run of the \textsc{jam} script resulted in a simulated $V_{rms}$ map. In order to determine the best values of the free parameters in the JAM modelling, we used Bayesian inference \citep{gel13}, assuming a constant prior and a Gaussian distribution of errors ($\propto$ $e^{-\chi^2/2}$), which, from the Bayes theorem, implies that the natural logarithm of the posterior is proportional to the $\chi^2$. The search for the best values of the free parameters in the JAM modelling was performed with an implementation of the Adaptive Metropolis (AdaMet - \citealt{haa01}) algorithm developed by \citet{cap13}. The posterior distributions resulting from this methodology provided not only the best values for the three free parameters considered in this modelling, but also its uncertainties. The $\chi^2$ values were calculated between the observed and simulated $V_{rms}$ maps, using the following equation:

\begin{equation}
\chi^2 = \frac{1}{I} \cdot \sum_{i=1}^{Nx} \sum_{j=1}^{Ny} \frac{I_{ij} \cdot (V_{rms~ij}(observed) - V_{rms~ij}(simulated))^2}{e_{Vij}^2},
\label{eq2}
\end{equation}

\noindent where $Nx$ and $Ny$ are the numbers of spaxels along the horizontal and vertical axes, respectively, $e_{Vij}$ is the uncertainty of $V_{rms}$ in the spaxel ($i$, $j$) of the observed $V_{rms}$ map, $V_{rms~ij}(observed)$ is the $V_{rms}$ value in the spaxel ($i$, $j$) of the observed $V_{rms}$ map, $V_{rms~ij}(simulated)$ is the $V_{rms}$ value in the spaxel ($i$, $j$) of the simulated $V_{rms}$ map, $I_{ij}$ is the value of the spaxel ($i$, $j$) of the image containing the integrated stellar fluxes of the corresponding data cube and $I$ is the sum of the values of all spaxels in the previous image. 

In the previous equation, we used the integrated stellar flux values of the data cubes ($I_{ij}$) as weight functions to calculate the $\chi^2$, in order to give more weight to the central areas (close to the SMBH). Similar approaches to obtain $\chi^2$ values were used in a number of previous studies (e.g. \citealt{pei03,men13,men15b}).

The simulated $V_{rms}$ maps corresponding to the best obtained models for NGC 3115 and NGC 4699 are shown in Figs.~\ref{fig2} and ~\ref{fig4}, respectively. The general morphologies of the observed $V_{rms}$ maps were well reproduced by all simulated maps. The extracted curves reveal that the simulated values were all compatible, at the 1-$\sigma$ or 2-$\sigma$ levels, with the observed ones. Although our dynamical modelling (and, more specifically, the calculation of the $\chi^2$ values) was based only on the $V_{rms}$ maps, the simulated $V_*$ maps corresponding to the best obtained models have also a significant importance in this work, as their shape reveal in detail the effects of our treatment methodology (in particular, the increase in the spatial resolution of the observations, provided by the Richardson-Lucy deconvolution). Therefore, in Figs.~\ref{fig3} and ~\ref{fig5}, we show the simulated $V_*$ maps, obtained with the same parameters determined from the analysis of the $V_{rms}$ maps, and assuming that the additional anisotropy parameter $\gamma = 1 - \sigma_\phi^2/\sigma_R^2$ = 0. Again, the spatial morphologies of the observed $V_*$ maps were well reproduced by the simulated maps. The simulated values in the extracted curves were all compatible, at the 1-$\sigma$ or 2-$\sigma$ levels, with the observed ones. The posterior distributions of the dynamical modellings applied to the non-treated, filtered and deconvolved data cubes of NGC 3115 and NGC 4699, are shown in Figs.~\ref{fig9} and~\ref{fig10}.

Tables~\ref{tbl2} and~\ref{tbl3} show the $M_{BH}$, $M/L_{I}$ and $\beta$ values, with the corresponding uncertainties, of the best obtained models for NGC 3115 and NGC 4699. For both galaxies, the values of each parameter derived from the non-treated, filtered and deconvolved data cubes are all compatible, at the 1-$\sigma$ level. On the other hand, it is clear that the Butterworth filtering decreased the uncertainties of all parameters and, then, the additional use of the Richardson-Lucy deconvolution decreased even more the uncertainties of the parameters. 

\section{Discussion}

\begin{table*}
\centering
\caption{Parameters of the best obtained models for the non-treated, filtered and deconvolved data cubes of NGC 3115. \label{tbl2}}
\begin{tabular}{cccc}
\hline
    & $M_{BH}$ ($10^8$ $M_{\sun}$) & $M/L_I$ ($M_{\sun}$/$L_{\sun I}$) & $\beta$ \\
\\ \hline
Non-treated & $4.8 \pm 2.9$ & $2.9 \pm 0.4$ & $-0.20 \pm 0.14$\\
Filtered & $3.9 \pm 2.1$ & $3.2 \pm 0.3$ & $-0.11 \pm 0.09$ \\
Deconvolved & $4.4 \pm 1.7$ & $3.15 \pm 0.23$ & $-0.13 \pm 0.07$ \\   
\hline
\end{tabular}
\end{table*}

\begin{table*}
\centering
\caption{Parameters of the best obtained models for the non-treated, filtered and deconvolved data cubes of NGC 4699. \label{tbl3}}
\begin{tabular}{cccc}
\hline
    & $M_{BH}$ ($10^8$ $M_{\sun}$) & $M/L_I$ ($M_{\sun}$/$L_{\sun I}$) & $\beta$ \\
\\ \hline
Non-treated & $3.2 \pm 2.2$ & $3.3 \pm 0.4$ & $-0.05 \pm 0.19$\\
Filtered & $2.9 \pm 1.5$ & $3.39 \pm 0.24$ & $-0.03 \pm 0.12$ \\
Deconvolved & $2.9 \pm 1.2$ & $3.37 \pm 0.20$ & $-0.03 \pm 0.09$ \\   
\hline
\end{tabular}
\end{table*}

The results shown in the previous section, particularly in Tables~\ref{tbl2} and~\ref{tbl3}, clearly indicate that the treatment procedure presented in \citet{men14a, men15a, men19} not only does not compromise analyses of data cubes focused on the stellar kinematics, but actually improves the quality of the results, in this case, by decreasing the uncertainties of the parameters provided by the modelling of stellar kinematics in the data cubes of NGC 3115 and NGC 4699. For $M_{BH}$, the Butterworth spatial filtering decreased the uncertainty in 28\% for NGC 3115 and 32\% for NGC 4699. For $M/L_I$, this decrease was of 25\% for NGC 3115 and 40\% for NGC 4699. For $\beta$, the decrease was 36\% for NGC 3115 and 37\% for NGC 4699. The use of the Richardson-Lucy deconvolution, after the Butterworth spatial filtering, resulted, for $M_{BH}$, in an additional decrease of the uncertainty of 19\% for NGC 3115 and 20\% for NGC 4699. For $M/L_I$, this additional decrease was of 23\% for NGC 3115 and 17\% for NGC 4699. For $\beta$, the decrease was of 22\% for NGC 3115 and 25\% for NGC 4699. If we consider the entire data treatment procedure (Butterworth spatial filtering together with Richardson-Lucy deconvolution), for $M_{BH}$, the decrease in the uncertainty was of 41\% for NGC 3115 and 45\% for NGC 4699. For $M/L_I$, this entire decrease was of 42\% for NGC 3115 and 50\% for NGC 4699. For $\beta$ the decrease was of 50\% for NGC 3115 and 53\% for NGC 4699.

The $M_{BH}$ value we obtained in this work for NGC 3115 is somewhat lower than the values of $\sim10^9$ M$_{\sun}$ determined by \citet{kor92}, \citet{kor96} and \citet{ems99}. This discrepancy could, in principle, be due to the fact that the data used in this work has a spatial resolution significantly higher than that of the data used in these previous works. On the other hand, our $M_{BH}$ value of NGC 3115 is compatible, at the 1-$\sigma$ level, with the estimate obtained by \citet{mag98} ($4.06^{+0.19}_{-0.22} \times 10^8$ $M_{\sun}$).

An interesting feature of the stellar kinematics in the central region of NGC 3115, which is easier to detect in the extracted $V_{rms}$ curves corresponding to the filtered and deconvolved data cubes in Fig.~\ref{fig2}, is that the observed $V_{rms}$ peak does not coincide with the stellar bulge centre, corresponding to the same position of the simulated $V_{rms}$ peak. In \citet{men14b}, we had already identified a displacement of the $\sigma_*$ peak relative to the stellar bulge centre, although no detailed discussion was provided. As mentioned in Section~\ref{sec1}, in \citet{men14b}, we reported the detection of an off-centred AGN (with a broad H$\alpha$ component) in NGC 3115, at a projected distance of $\sim0.29 \pm 0.05$ arcsec (corresponding to $\sim14.3 \pm 2.5$ pc) from the stellar bulge centre. This off-centred AGN could, in principle, be part of a SMBH binary system (the other SMBH being located probably at the stellar bulge centre) or could be a ``kicked'' SMBH after the coalescence of two SMBHs. Based on the findings of \citet{jon19}, who concluded that the position of the radio source in this galaxy is the same of the stellar bulge centre (indicating the presence of a probable AGN there), the scenario involving a SMBH binary system is the most likely. The impact of such a binary system on the stellar kinematics in the nuclear region of this galaxy could explain the observed displacement of the $V_{rms}$ peak relative to the stellar bulge centre.

As discussed in Section~\ref{sec1}, there is not much information in the literature about the central region of NGC 4699. Using maximum entropy dynamical models, \citet{bow93} established an upper limit of $3 \times 10^9$ M$_{\sun}$ for the mass of this SMBH, which is considerably higher than the value we obtained in this work. It is difficult to establish a comparison with this result determined by \citet{bow93}, as it is just an upper limit for the SMBH mass and the data we used in this work has a higher spatial resolution. On the other hand, our $M_{BH}$ values are compatible, at the 1-$\sigma$ level, with the result of \citet{erw15} and \citet{sag16} ($(1.76 \pm 0.23) \times 10^8$ M$_{\sun}$) from the modelling of the stellar kinematics in AO SINFONI data. The lower $M_{BH}$ uncertainty of the value determined by \citet{erw15} and \citet{sag16} is certainly a consequence of the high spatial resolution of the SINFONI data. It is worth emphasizing that the data treatment procedure we discussed in this work can also be applied to AO data (see \citealt{men14a,men15a}), which would probably result in even lower uncertainties for $M_{BH}$.

Besides comparing the values of $M_{BH}$ we obtained with those determined by previous studies, another way to check the reliability of the dynamical modelling we applied involves the evaluation of the $M/L_I$ values provided by the models. If we take, for example, the $M/L_I$ ratio vs $\sigma_e$ (the stellar velocity dispersion at the effective radius of the galaxies) correlation obtained by \citet{cap06} (equation 7 in their paper), based on data of the Spectroscopic Areal Unit for Research on Optical Nebulae survey \citep{bac01}, we obtain, for NGC 3115, assuming $\sigma_e = 230 \pm 11$ km s$^{-1}$ \citep{gul09}, $M/L_I$ = $4.27 \pm 0.24$ $M_{\sun}$/$L_{\sun I}$, which is compatible, at the 3-$\sigma$ level, with the values derived from the non-treated and deconvolved data cubes and, at the 2-$\sigma$ level, with the value determined from the filtered data cube. If we use this same correlation for NGC 4699, assuming a central $\sigma$ of $193 \pm 9$ km s$^{-1}$ (corresponding to the most reliable value we found, from Hyperleda), we obtain $M/L_I$ = $3.69 \pm 0.20$ $M_{\sun}$/$L_{\sun I}$, compatible, at the 1-$\sigma$ level, with all the values derived from our models. 

We can also compare the $M_{BH}$ we obtained with those derived from the $M_{BH}$ - $\sigma$ relation. By using the version of this relation given by \citet{kor13} (equation 7 in their review) and $\sigma_e = 230 \pm 11$ km s$^{-1}$, we obtain $M_{BH} = (5.7 \pm 1.4) \times 10^8$ M$_{\sun}$ for NGC 3115, which is compatible, at the 1-$\sigma$ level, with the values obtained in this work. Applying the same procedure to NGC 4699, using a central $\sigma$ of $193 \pm 9$ km s$^{-1}$, we obtain $M_{BH} = (2.6 \pm 0.6) \times 10^8$ M$_{\sun}$, again compatible, at the 1$\sigma$ level, with our results.  

The existence of a correlation between the masses of the central SMBHs in galaxies and the stellar velocity dispersion of the bulges suggests a possible co-evolution between the SMBHs and their host galaxies. In that scenario, AGNs feedback, regulating, for example, the star formation in the host galaxies, certainly plays a significant role. The $M_{BH}$ - $\sigma$ relation shows an intrinsic scatter. At the high-mass end, where the most massive elliptical galaxies are located, there is less scatter. On the other hand, at the low-mass end, a SMBH of a given mass can be found in variety of galaxy masses, which results in a larger scatter in the relation. The evaluation of the intrinsic scatter of the $M_{BH}$ - $\sigma$ relation has a significant impact on theories of formation and evolution of galaxies, on the estimate of the space density of SMBHs and on studies of the evolution of the $M_{BH}$ - $\sigma$ relation in time (for further detail, see \citealt{gul09}). One of the main difficulties to perform an analysis of the intrinsic scatter in the $M_{BH}$ - $\sigma$ relation is to separate it from the observational scatter. For that purpose, a reduction in the uncertainties of the $M_{BH}$ values can be very relevant. That is an example of a very important use of the treatment procedure we discussed in this work. It is worth mentioning, however, that, as can be seen by the improvements in the observed kinematic maps presented in Section~\ref{sec31}, provided by the treatment procedure, this methodology can also be very useful for other analyses focused on the stellar or even gas kinematics in data cubes.

\section{Conclusion}

We evaluated the efficacy of our data cube treatment methodology, described in \citet{men14a,men15a,men19}, for analyses focused on the stellar kinematics. To do that, we used GMOS/IFU data cubes of the central regions of the galaxies NGC 3115 and NGC 4699 and applied a dynamical modelling with Jeans Anysotropic Models. The main conclusions of this work are:

\begin{itemize}

\item Both for NGC 3115 and NGC 4699, all the values of $M_{BH}$, $M/L_I$ and $\beta$, from the non-treated, filtered and deconvolved data cubes, were compatible, at the 1-$\sigma$ level. However, the use of the Butterworth spatial filtering reduced the uncertainty of the parameters. The additional use of the Richardson-Lucy deconvolution reduced even more the uncertainty of the parameters.

\item The complete data treatment procedure resulted in reductions of 41\% and 45\% of the uncertainties of $M_{BH}$ for NGC 3115 and NGC 4699, respectively. 

\item The $M_{BH}$ values we obtained for NGC 3115 ($(4.4 \pm 1.7) \times 10^8$ $M_{\sun}$ from the deconvolved data cube) and for NGC 4699 ($(2.9 \pm 1.2) \times 10^8$ $M_{\sun}$ from the deconvolved data cube) are compatible, at the 1-$\sigma$ level, with those derived from the $M_{BH}$ - $\sigma$ relation.

\item The analysis of NGC 3115 revealed that the observed $V_{rms}$ peak is displaced from the stellar bulge centre, which is consistent with findings of previous studies. Considering that and the previous detection of an off-centred AGN in this galaxy, we conclude that a scenario involving a SMBH binary system, with one SMBH at the stellar bulge centre and the other located at a projected distance of $\sim0.29 \pm 0.05$ arcsec (corresponding to $\sim14.3 \pm 2.5$ pc) from the centre, is the most likely to explain the off-centred $V_{rms}$ peak. 

\item The results of the dynamical modelling clearly indicate that our treatment procedure not only does not compromise analyses of data cubes focused on the stellar or gas kinematics, but actually improves the quality of the results.

\end{itemize}

\section*{Acknowledgements}

This research was based on observations made with the NASA/ESA \textit{HST} obtained from the Space Telescope Science Institute, which is operated by the Association of Universities for Research in Astronomy, Inc., under NASA contract NAS 5–26555. These observations are associated with programmes 5512 and 15133. This work was based on observations obtained at the Gemini Observatory (processed using the Gemini \textsc{iraf} package), which is operated by the Association of Universities for Research in Astronomy, Inc., under a cooperative agreement with the NSF on behalf of the Gemini partnership: the National Science Foundation (United States), the National Research Council (Canada), CONICYT (Chile), the Australian Research Council (Australia), Minist\'erio da Ci\^encia, Tecnologia e Inova\c{c}\~ao (Brazil) and Ministerio de Ciencia, Tecnolog\'ia e Innovaci\'on Productiva (Argentina). We also acknowledge the usage of the HyperLeda data base (http://leda.univ-lyon1.fr). This research has made use of the NASA/IPAC Extragalactic Database (NED), which is operated by the Jet Propulsion Laboratory, California Institute of Technology, under contract with the National Aeronautics and Space Administration. PS acknowledges Funda\c{c}\~ao de Amparo \`a Pesquisa do Estado de S\~ao Paulo (FAPESP) post-doctoral fellowship 2020/13239-5. RBM and TVR thanks Conselho Nacional de Desenvolvimento Cient\'ifico e Tecnol\'ogico (CNPq) for the support under grants 309976/2022-7 and 304584/2022-3, respectively.

\section*{Data Availability}

The data used in this work is from the DIVING$^{3D}$ survey, obtained with Gemini-South telescope. All raw data are publicly available in the archive of the Gemini Observatory. The treated data can be acquire under request to diving3d@gmail.com.



\bibliographystyle{mnras}
\bibliography{reference} 



\bsp	
\label{lastpage}
\end{document}